\documentclass{JHEP3}

\usepackage{graphicx}
\usepackage{latexsym}
\usepackage{amsmath}
\usepackage{amsfonts}
\usepackage{amssymb}

\newcommand{\be}{\begin{equation}}
\newcommand{\ee}{\end{equation}}
\newcommand{\ben}{\begin{eqnarray}}
\newcommand{\een}{\end{eqnarray}}

\title{Tachyon condensation on brane sphalerons}

\author{Francisco A. Brito
\\
Departamento de F\'\i sica, Universidade Federal de Campina
Grande,\\
58109-970 Campina Grande, Para\'\i ba, Brazil\\
E-mail: fabrito@df.ufcg.edu.br}

\abstract{We consider a sphaleron solution in field theory that
provides a toy model for unstable $D$-branes of string theory. We
investigate the tachyon condensation on a $Dp$-brane. The localized
modes, including a tachyon, arise in the spectrum of a sphaleron
solution of a $\phi^4$ field theory on ${\mathbb{M}\,}^{p+1}\times
S^1$. We use these modes to find a multiscalar tachyon potential
living on the sphaleron world-volume. A complete cancelation between
brane tension and the minimum of the tachyon potential is found as
the size of the circle becomes small.\\

\vspace{1cm}

}

\maketitle

\begin{document}

\section{Introduction}

The existence of tachyon modes living on a D-brane anti-D-brane pair
\cite{banks,green} or on a non-BPS D-brane
\cite{sen5,bergman,sen9,sen10,witten98,horava,harvey12} of type IIA
or IIB string theory is associated with the spectrum of open strings
ending on D-branes. Several general arguments
\cite{sen3,sen4,sen5,ot1,ot2,ot3,ot4} assert the tachyon potential
has a minimum that represents the closed string vacuum without
D-branes. In order for this to be true the negative energy density
given by the tachyon potential at the minimum must exactly cancel
the sum of the tensions of a D-brane anti-D-brane pair or the
tension of a non-BPS D-brane (Sen's conjecture). The evidence of
this conjecture by using string field theory \cite{r1} has been well
explored in the literature
\cite{r2,r3,r4,r5,r6,r7,r8,r9,r10,r11,r12,r13,r14,r15}. Following
the level truncation scheme initiated in Ref.~\cite{kosteleck},
specially in Refs.~\cite{sen_zwiebach,berko_sen_zwie}, the evidence
of this conjecture was first investigated in open bosonic string
field theory \cite{sen_zwiebach} and in open superstring field
theory \cite{berko_sen_zwie}, where the vacuum of the tachyon
potential cancels about $99\%$ and $85\%$ of the D-brane tension,
respectively. This is a phenomenon of tachyon condensation whose
analysis of course requires computation of the tachyon potential.
For earlier developments on tachyon condensation see
Refs.~\cite{halpern}. While such computation in string theory is
complicated, in field theory the computation of the tachyon
potential is much simpler. Thus, a way of investigating the tachyon
condensation involving analytical solutions is the use of toy models
in field theory. An interesting study on this direction can be found
in Refs.~\cite{zwiebach1,kraus}.

In this paper, following the lines of \cite{zwiebach1}, we present
another example where we can easily investigate the tachyon
condensation. We study a field theory presenting a sphaleron
solution whose discrete spectrum has a tachyon mode, the zero mode
and other massive modes. We integrated out these modes over a
compact coordinate to find an effective action that we identify as
describing the world-volume of a non-BPS $Dp$-brane of string field
theory. We focus our attention on the multiscalar tachyon potential
living on this world-volume to study tachyon condensation at several
levels.

Sphaleron solutions are static and unstable classical solutions,
localized in space, whose existence is associated with
non-contractible loops in the space of field configurations
\cite{manton83,manton84}. In string theory unstable D-branes can
be understood as analogs of sphalerons of field theory, and can be
referred to as D-brane sphalerons or simply ``D-sphalerons''
\cite{harvey12}. Furthermore the AdS/CFT correspondence has been
used to show that unstable $D0$-brane in type IIB on $AdS_5$ has a
field theory dual which is a sphaleron in gauge theories on
$S^3\times \mathbb{R}$ \cite{gross}--- see also \cite{kars} for
recent developments. Following the perspective of
Ref.~\cite{gross}, one could investigate other field theoretical
models, exhibiting sphalerons, being corresponding holographic of
strings moving on a higher dimensional special manifold ${\cal
M}$, just as in the AdS/CFT correspondence. For simplicity, in
this paper, we focus only on the study of sphaleron solution of a
$\phi^4$ theory on ${\mathbb{M}\,}^{p+1}\times S^1$, where
${\mathbb{M}\,}^{p+1}$ is a $p+1$ dimensional Minkowski space. One
can regard this field theory in ${\mathbb{M}\,}^{p+1}\times S^1$
as a corresponding holographic of strings moving on a special
manifold ${\cal M}_{p+3}$. In this picture, one may find that
sphalerons on the field theory side are dual to D-sphalerons
\cite{harvey12} on the string theory side. This $\phi^4$ theory,
up to a constant, is the open superstring field
\cite{berko_sen_zwie,irina} restricted to the tachyon only
\cite{zwiebach1,kraus}. The sphaleron solution that we are dealing
with is a static unstable periodic solution given in terms of
Jacobi elliptic function. As one knows this solution is labeled by
a real elliptic modulus parameter $0\leq k<1$
\cite{arscott,whitaker}. We show that as $k$ approaches zero the
size of the circle $S^1$ becomes small and the cancelation between
the negative energy density contribution from the tachyon
potential and the tension of the non-BPS $Dp$-brane becomes more
and more accurate at lower levels. In the limit $k\to0$ the
multiscalar tachyon potential we obtain is exact at level zero,
being the extra scalars just spurious states. This tachyon
potential has two minima that exactly cancels the tension of the
non-BPS $Dp$-brane. This result in field theory mimics the
expected result in string theory according to Sen's conjecture.
Since the minima of the tachyon potential are invariant under
$Z_2$ symmetry there exists tachyon kink (anti-kink) connecting
these minima representing a $D$-brane with one lower dimension
\cite{asad_terashina,berko_sen_zwie,bbn}. The paper is organized
as follows. In Sec.~\ref{action} we obtain the spectrum of the
sphaleron solution and the effective action. In Sec.~\ref{eff_pot}
we obtain the multiscalar tachyon potential and discuss the
tachyon condensation. In Sec.~\ref{conclu} we present our comments
and conclusions.

\section{Effective action for the sphaleron solution spectrum}
\label{action}

In this section we consider a scalar field theory in $p+2$
dimensional space-time with the topology $\mathbb{M\,}^{p+1}\times
S^1$. The action is \ben \label{spha}S&=&\int
dt\,d^py\,dx\left[\frac{1}{2}\partial_M\Phi\,\partial^M\Phi-V(\Phi)\right],
\nonumber\\
&=&\int
dy^{p+1}\,dx\left\{\frac{1}{2}\left[\left(\frac{\partial\Phi}{\partial
t}\right)^2-\left(\frac{\partial\Phi}{\partial
y_i}\right)^2-\left(\frac{\partial\Phi}{\partial x}\right)^2\right]
-V(\Phi) \right\},\een with $M=(y^\mu,x)$ and $\Phi$ is a real
scalar field. $y^\mu=(t,y_i)$, $i=1,2,...,p$, are coordinates on the
world-volume of a $Dp$-brane embedded in a $p+2$ dimensional
space-time with the coordinate $x$ compactified on a circle $S^1$.
This $Dp$-brane is represented by a static nontopological soliton
solution $\bar{\Phi}$ of the equation of motion \ben \label{eom}
\square_{(p+2)}{{\Phi}}=-\frac{\partial V}{\partial\Phi}, \een for
$\bar{\Phi}$ depending only on the compact coordinate $x$. We take
the positive semi-definite scalar potential of a $\Phi^4$ theory
\ben \label{Vphi} V(\Phi)=\frac{1}{2}\,(\Phi^2-1)^2. \een This
potential, up to a constant, can be viewed as the truncation of open
superstring field theory \cite{berko_sen_zwie,irina} restricted to
the tachyon field only \cite{zwiebach1,kraus}. In the background
(\ref{spha}), this potential in the unstable vacuum $\Phi=0$, i.e.,
before tachyon condensation, is the tension of the space-filling
$(p+1)$-brane given by $T_{p+1}=V(\Phi=0)$. Below we study the
tachyon condensation from the point of view of a field theory living
on unstable solution of this theory. This solution defines an
unstable $Dp$-brane with tension $T_p$.

As is well-known the potential (\ref{Vphi}) does not produce lump
solutions with tachyon mode on its fluctuation spectrum when the
coordinate $x$ has infinite size. However when $x$ is compactified
on a circle of length $L$ \cite{manton,brihaye,liang,bbb} this
potential does provide static periodic solutions
$\bar{\Phi}(x)=\bar{\Phi}(x+L)$, with tachyon mode on its
fluctuation spectrum. They are sphaleron solutions and can be
expressed in terms of Jacobi elliptic function, i.e., \ben
\bar{\Phi}(x)=k\,b(k)\,{\rm sn}{(b(k)\,x,k)} \label{spha_sol},\qquad
b(k)=\sqrt{\frac{2}{1+k^2}}, \een where $0\leq k<1$ is a real
elliptic modulus parameter. This is a static periodic function with
period $4K(k)/b$, where $K(0)=\pi/2$ and $K(1)=\infty$. $K(k)$ is
the complete elliptic integral of the first kind. Note that for a
coordinate $x$ with infinite size, the period is infinite ($k=1$)
and the solution becomes $\bar{\Phi}(x)=\tanh{(x)}$, which is a kink
solution without any tachyon mode on its spectrum. Let us now
observe the following. In conformal field theory, e.g. four
dimensional Yang-Mills theory, there are no static finite energy
solution (stable or unstable) because there is no scale to fix the
mass of the solution. However sphalerons can be found if one
considers the theory on $S^3\times \mathbb{R}$ \cite{gross}. The
size $R$ of the sphere $S^3$ fixes the mass of any static solution
as $\sim1/R$. Similarly, in $\Phi^4$ theory, the sphaleron solution
can be found only if a coordinate is compactified on a circle $S^1$
with size $L=4K(k)/b$. Furthermore, the mass of the tachyon is fixed
as ${M_0}^2\sim-\,1/{R\,}^2$, where one defines $R=L/2\pi$. In the
limits discussed above, we readily find masses ${M_0}^2\to-\,2$
($k\to0$) and ${M_0}^2\to 0$ ($k\to1$), as we can confirm latter by
computing the sphaleron spectrum.

Let us now expand the action (\ref{spha}) around the sphaleron
solution (\ref{spha_sol}). We follow the lines of
Ref.~\cite{zwiebach1} performed for lump solution. Consider the
transformation $\Phi\to\bar{\Phi}+\eta$, $S(\Phi)\to
S(\bar{\Phi}+\eta)$ into (\ref{spha}) such that \ben\label{spha_eta}
S=\int d^{p+1}y\,dx\left[-\frac{1}{2}
\left(\frac{d\bar{\Phi}}{dx}\right)^2-V(\bar{\Phi})
-\frac{1}{2}\partial_\mu\eta\,\partial^\mu\eta
+\frac{1}{2}\left(\eta\,\frac{\partial^2\eta}{\partial x^2}-\eta
V{''}(\bar{\Phi})\,\eta\right)\right.\nonumber\\
\left.-\frac{V{'''}(\bar{\Phi})}{3!}\eta^3
-\frac{V{''''}(\bar{\Phi})}{4!}\eta^4\qquad\right],
 \een
(the primes mean derivatives with respect to argument of the
function.) Since we are considering the $\Phi^4$ theory
(\ref{Vphi}), the expansion above is exact. We define above
$y^\mu=(t,y_i)$ with a mostly plus signature $(-,+,+,...,+)$. The
first two terms of the expansion are related to the energy of the
sphaleron solution or simply $Dp$-brane tension \ben \label{eqTp}
T_p=\int_0^{4K/b} dx\left[\frac{1}{2}
\left(\frac{d\bar{\Phi}}{dx}\right)^2+V(\bar{\Phi})\right].\een The
fluctuations on the $Dp$-brane is governed by the quadratic $\eta$
terms of (\ref{spha_eta}). They provide a Schroedinger-like equation
for the fluctuations $\eta$ given as \ben \label{sch}
-\frac{d^2\psi_n(x)}{dx^2}+V{''}(\bar{\Phi}(x))\,\psi_n(x)=
{M_n}^2\,\psi_n(x),\een where we have considered the normal mode
expansion \ben\label{modes} \eta(y^\mu;x)=\sum_n\xi_n(y)\psi_n(x),
\een and the fact that the fields $\xi_n(y)$ living on the
$Dp$-brane world-volume satisfy their equations of motion
\ben\label{xin} \square_{(p+1)}\,\xi_n(y)={M_n}^2\xi_n(y). \een Let
us now use (\ref{sch}), the scalar potential given in (\ref{Vphi})
and the sphaleron solution (\ref{spha_sol}) to obtain its spectrum.
The Schroedinger-like equation now reads \ben \label{sch_sol}
-\frac{d^2\psi_n(x)}{dx^2}+[6\,k^2\,b(k)^2\,{\rm
sn}^2{(b(k)\,x,k)}-2]\,\psi_n(x)={M_n}^2\,\psi_n(x).\een This can be
recognized as Lam\'e equation \ben\label{lame}
-\frac{d^2\psi_n(z)}{dz^2}+N(N+1)\,k^2\,{\rm sn}^2{(z,k)}\,\psi_n(z)
=h_n\,\psi_n(z), \een with $z=b(k)\,x$ and
$h_n=({M_n}^2+2)/\,b(k)^2$. The spectrum of this quantum mechanics
problem is well-known \cite{arscott,whitaker}. The spectrum concerns
$2N+1$ discrete states (for $N$ positive integer) that are edges of
$N$ bound bands followed by a continuum band --- see also
\cite{khare,dunne}. We shall focus only on the band edges, i.e., the
discrete states. This will be further justified as we discuss the
level expansion in Sec.~\ref{eff_pot}. In the case we are
considering above $N\!=\!2$, so that we have five discrete states
with eigenfunctions given in terms of Jacobi elliptic functions and
respective eigenvalues \ben \label{spectrum} \psi_0&=&{\rm
sn}^2{(z,k)}-\frac{1+k^2+\sqrt{1-k^2+k^4}}{3k^2},\!\!\!\!\!\!\!\!\!\!\!\!
\qquad  {M_0}^2\!=\!(1+k^2-2\sqrt{1-k^2+k^4})\,b(k)^2,\\
\psi_1&=&{\rm cn}(z,k)\,{\rm
dn}(z,k),\!\!\!\!\!\!\!\!\!\!\!\!\qquad\qquad \qquad\qquad
\qquad {M_1}^2=0,\\
\psi_2&=&{\rm sn}(z,k)\,{\rm
dn}(z,k),\!\!\!\!\!\!\!\!\!\!\!\!\qquad\qquad \qquad\qquad \qquad
{M_2}^2=
3\,k^2\,b(k)^2, \\
\psi_3&=&{\rm sn}(z,k)\,{\rm
cn}(z,k),\!\!\!\!\!\!\!\!\!\!\!\!\qquad\qquad \qquad\qquad
\qquad {M_3}^2=3\,b(k)^2,\\
\label{spectrum4} \psi_4&=&{\rm
sn}^2{(z,k)}-\frac{1+k^2-\sqrt{1-k^2+k^4}}{3k^2},
\!\!\!\!\!\!\!\!\!\!\!\!\qquad
{M_4}^2\!=\!(1+k^2+2\sqrt{1-k^2+k^4})\,b(k)^2.\een Note that in the
interval $0\leq k<1$ the mass ${M_0}^2<0$, which implies that
$\psi_0$ is always a tachyon mode. We have the zero mode $\psi_1$
and the remaining are all massive modes. Now we are ready to compute
the action of the $Dp$-brane world-volume. Let us restrict ourselves
to discrete modes only, i.e., \ben\label{modes_exp}
\eta(y^\mu;x)=\xi_0(y)\psi_0(x)+\xi_1(y)\psi_1(x)+\xi_2(y)\psi_2(x)
+\xi_3(y)\psi_3(x)+\xi_4(y)\psi_4(x). \een  We substitute this mode
expansion into Eq.~(\ref{spha_eta}) and integrate out all the modes
over a period $4K(k)/b$. We have normalized the eigenfunctions in
this period and used the orthonormality condition \ben
\int_0^{\,4K/b}dx\,{\psi_n(x)\,\psi_m(x)}=\delta_{m,n}. \een This
gives rise to a field theory living on the world-volume of the
sphaleron representing a non-BPS $Dp$-brane with action
\ben\label{eff_actionS} S_p=\int
d^{p+1}y\left\{-T_p-\frac{1}{2}\sum_{n=0}^4\partial_\mu\,\xi_n(y)
\,\partial^\mu\,\xi_n(y)-V(\xi)\right\}. \een This is a theory of
five real scalar fields in $p+1$ dimensional space-time that we get
from a $\Phi^4$ theory living in $p+2$ dimensions as we compactify
one extra spatial dimension. The term $T_p$ is the tension of the
$Dp$-brane given by (\ref{eqTp}). In the tachyon condensation, this
term must exactly cancel the minima of the tachyon potential at
critical points $\xi^*$, i.e., \ben\label{conject} T_p+V(\xi^*)=0.
\een At $\xi=\xi^*$ the non-BPS $Dp$-brane is indistinguishable from
the vacuum where there is no $D$-branes. This is analogous to Sen's
conjecture in string theory \cite{r14}.

One can also translate the action (\ref{eff_actionS}) to the usual
DBI like action for the tachyon field dynamics on the world-volume
of a $Dp$-brane in string theory \cite{garousi,sen0204143}, i.e.,
\ben\label{stringTach}
S_p=-\int{d^{p+1}y\,V(T)\sqrt{1+\partial_\mu T\partial^\mu T}}\,.
\een Let us consider the action (\ref{eff_actionS}) truncated up
to the tachyon field $\xi_0$, and assume the following procedure
\cite{minahan,bbn}:
\ben\label{VTxi}&&V(\xi_0)=\left[\frac{\partial\xi_0(y)}{\partial
T}\right]^2-T_p=V(T)-T_p\,,\\
&&\partial_\mu\xi_0(T(y))=\frac{\partial\xi_0(y)}{\partial T}\,
\partial_\mu T(\xi_0(y)).\een
Now upon such considerations we can write \ben \label{xi0Tach}
S_p&=&\int d^{p+1}y\left\{-T_p-\frac{1}{2}\partial_\mu\,\xi_0(y)
\,\partial^\mu\,\xi_0(y)-V(\xi)\right\}\\
&=&\int d^{p+1}y\left\{-V(T)-\frac{1}{2}V(T)\,\partial_\mu
T\partial^\mu T \right\}. \een This is precisely the action
(\ref{stringTach}) expanded up to quadratic first derivative
terms. Such an approximation is good as long as higher order
derivatives of $T$ are small. This is indeed the case for tachyon
matter \cite{sen0204143}. Finally, the inclusion of the other
scalar fields becomes straightforward as we use the same
procedure.

\section{The multiscalar tachyon potential and tachyon condensation}
\label{eff_pot}

As we mentioned above we should integrate out all the modes over
the compact coordinate $x$ into the action (\ref{spha_eta}) in
order to find the effective action of the modes living on the
world-volume of the sphaleron (\ref{eff_actionS}). The multiscalar
tachyon potential living on such world-volume reads \ben
V(\xi)=\int_0^{4K/b} dx\left\{
\frac{1}{2}\left(\eta\,\frac{\partial^2\eta}{\partial x^2}-\eta
V{''}(\bar{\Phi})\,\eta\right)
-\frac{V{'''}(\bar{\Phi})}{3!}\eta^3
-\frac{V{''''}(\bar{\Phi})}{4!}\eta^4\right\}.\een This is an
integration involving the sphaleron solution $\bar{\Phi}$ and the
normal modes $\psi_n$ performed in a period $4K/b$. Since these
objects are labeled by the parameter $0\leq k<1$, this means we
have distinct multiscalar potentials for distinct values of $k$.

Let us now analyze the tachyon condensation in the level
expansion. We test the validity of the equation (\ref{conject}),
i.e., we investigate how much $V(\xi^*)$ approaches the tension
$T_p$ at each level. We do this for several values of $k$. As is
usual we assign {\it level zero} to the tachyon field $\xi_0$. To
any other field $\xi_i$ ($i=1,2,3,4$) we must assign the {\it
level} $L_i=|{M_i}^2-{M_0}^2|$, with masses ${M_i}^2$ given by
equations (\ref{spectrum})-(\ref{spectrum4}). In the limit
$k\to0$, the level of each field becomes $L_1\to2$, $L_2\to2$,
$L_3\to8$ and $L_4\to8$, while in the limit $k\to1$ we find
$L_1\to0$, $L_2\to3$, $L_3\to3$ and $L_4\to4$. This means that for
small values of $k$ the fields become strongly effective in
contributing to the potential even in lower levels, while for
$k\to1$ the fields become weakly effective to make such
contribution so that we need more and more fields, i.e., higher
levels. In other words, the multiscalar tachyon potential at the
critical points, i.e., $V(\xi^*)$ approaches the tension $T_p$ as
we add scalar fields. This approach happens very quickly for
$k\to0$ and very slowly for $k\to1$. As a consequence, in the case
$k\to0$ few scalar fields are relevant while in the opposite case
$k\to1$ most of the scalar fields, including those from the
continuum spectrum, become relevant. This is evident from the
explicit calculations and Fig.~\ref{fig01} below. Thus it is
reasonable to expect that in the regime $k\to1$, the contributions
to the potential coming from continuum states become important.
These are states of the bound and continuum bands we mentioned
earlier. For each $k<1$ large enough, there are zones where the
values of energy allowed form a continuum band (Brillouin zone).
Below, however, we focus our attention on values of $k>0$, but
small enough.

We now first focus our attention on the multiscalar tachyon
potential living on the sphaleron world-volume for $k^2=1/32$. The
potential is given in terms of five scalar fields \ben
\label{k_32th} V(\xi)&=& - { \frac {232}{255}} \,\xi_0^{2}+{ \frac
{1}{11}} \,\xi_2^{2} + { \frac {32}{11}} \,\xi_3^{2} + { \frac
{355}{122}} \,\xi_4^{2} + { \frac {200}{1817}} \,\xi_0^{4} + { \frac
{194}{1167}} \,\xi_1^{4}+ { \frac { 179}{1094}} \,\xi_2^{4}+ { \frac
{63}{382}} \,\xi_3^{4} + { \frac {145}{ 879}} \,\xi_4^{4}\nonumber\\
\mbox{}&&+{ \frac {285}{302}} \,\xi_1^{2}\,\xi_4 \,\xi_0 - {\frac
{270}{293}} \,\xi_4\,\xi_2^{2}\, \xi_0 - { \frac {9}{1216}}
\,\xi_3^{2}\,\xi_4\,\xi_0 - { \frac {150}{217}}
\,\xi_3\,\xi_1\,\xi_0 - { \frac {170}{247}}
\,\xi_2\,\xi_4\,\xi_0 + { \frac {47}{3176}} \,\xi_0^{3}\,\xi_4  \nonumber \\
\mbox{}&& + { \frac {105}{163}} \,\xi_2^{2}\,\xi_0^{2}
 + { \frac {177}{262}} \,\xi_1^{2}\,\xi_0^{2} +
{ \frac {162}{325}} \,\xi_4^{2}\,\xi_2 + { \frac {42}{2675}}
\,\xi_2\,\xi_3\,\xi_4\,\xi_1 - { \frac {334}{179}}
\,\xi_1\,\xi_3\,\xi_2\,\xi_0 - { \frac {16}{2163}}
\,\xi_4^{3}\,\xi_0 \nonumber\\
\mbox{}&& + { \frac {188}{285}} \,\xi_0^{2}\,\xi_3^{2} + { \frac
{118}{179}} \,\xi_0^{2}\,\xi_4^{2} + { \frac {191}{579}}
\,\xi_1^{2}\,\xi_2^{2} + { \frac {239}{363}} \,\xi_3^{2}\,\xi_1^{2}
 + { \frac {173}{266}} \,\xi_1^{2}\,\xi_4^{2} + { \frac {39}{59}} \,\xi_3^{2}\,\xi
2^{2} + { \frac { 220}{329}} \,\xi_2^{2}\,\xi_4^{2}  \nonumber\\
\mbox{}&& + { \frac {223}{676}} \,\xi_4^{2}\,\xi_3^{2}
 + { \frac {1161}{2423}} \,\xi_0^{2}\,\xi_2 +
{ \frac {208}{851}} \,\xi_1^{2}\,\xi_2 + { \frac {41}{5286}}
\,\xi_3\,\xi_4\,\xi_1 + { \frac {166}{681}} \,\xi_2^{3} + { \frac
{239}{487}} \,\xi_3^{2}\,\xi_2 .\een As a first approximation let us
consider the multiscalar tachyon potential at level zero where only
the tachyon field $\xi_0$ is present, i.e., \ben \label{t0}
V_0=-\frac{232}{255}\,\xi_0^2+\frac{200}{1817}\,\xi_0^4. \een The
nontrivial critical points are $\xi^*_0\simeq\pm\,2.03293$ in which
the potential assumes the absolute value
$|V^{0}(\xi^*)|\simeq1.88001$. The $Dp$-brane tension is
$T_p\simeq2.27068$, thus \ben \label{ratio_32th}
\frac{|V^{0}(\xi^*)|}{T_p}\simeq0.827950, \een which corresponds to
about $82.79\%$ of the expected answer. Before going to the next
level let us study the expectation value of the fields at zero
energy of vacuum \ben \label{vev}<\xi_n>=\int_0^{\,4K/b}
dx\,\psi_n(x)\,(\Phi_0-\bar{\Phi}(x)). \een We define the critical
point of the tachyon potential as \cite{zwiebach1}
\ben\Phi_0-\bar{\Phi}(x)=\xi_0(y)\psi_0(x)+\xi_1(y)\psi_1(x)+\xi_2(y)\psi_2(x)
+\xi_3(y)\psi_3(x)+\xi_4(y)\psi_4(x),\een  being $\Phi_0=1$, the
vacuum of original ${\Phi\,}^4$ theory (\ref{Vphi}). After computing
(\ref{vev}) for $k^2=1/32$, we find that the expectation values
$<\xi_1>$ and $<\xi_3>$ are zero. Thus these fields play no role in
our analysis of tachyon condensation, such that they can be removed
from the theory. These are analogs of twist odd states in string
field theory \cite{zwiebach1}. The level of each remaining scalar
field $\xi_2$ and $\xi_4$ is $L_2\simeq2$ and $L_4\simeq8$,
respectively. So we can work out approximations $(L,I)$ with fields
up to level $L$ involving interactions up to level $I$.

Let us first consider the approximation $(2,8)$, where terms
involving the field $\xi_2$ up to fourth power are allowed. In
this approximation the multiscalar tachyon potential reads \ben
\label{2_8}
V^{(2,8)}=-\frac{232}{255}\,\xi_0^2+\frac{1}{11}\,\xi_2^2
+\frac{166}{681}\,\xi_2^3+ \frac{1161}{2423}\,\xi_0^2\,\xi_2+
\frac{105}{163}\,\xi_0^2\,\xi_2^2
+\frac{200}{1817}\,\xi_0^4+\frac{179}{1094}\,\xi_2^4. \een This
potential has the nontrivial critical points
$\xi^*_0\simeq\pm2.12944,\; \xi^*_2\simeq-0.373448$, where
$|V^{(2,8)}(\xi^*)|\simeq2.26312$. At this level we find
\ben\frac{|V^{(2,8)}(\xi^*)|}{T_p}\simeq0.996671,\een that
corresponds to about $99.67\%$ of the expected result. This nicely
improve the result obtained at level zero. We can still proceed up
to approximation (8,32), where we turn on the field $\xi_4$ to
which we assign the level $L_4\simeq8$, and consider its
interactions up to fourth power. Now the critical points of the
potential are \ben \label{cr_32th}\xi^*_0 \simeq -2.13221,\:
\xi^*_2 \simeq -0.37193,\: \xi^*_4 \simeq 0.03589.\een They are in
perfect agreement with expectation values $<\!\xi_0\!>$,
$<\!\xi_2\!>$, and $<\!\xi_4\!>$, as we can check by using
(\ref{vev}). At these critical points
$|V^{(8,32)}(\xi^*)|\simeq2.27062$ and then\ben
\label{ratio_32th_01234}
\frac{|V^{(8,32)}(\xi^*)|}{T_p}\simeq0.999974. \een In this
approximation the cancelation between the brane tension and the
minimum of the tachyon potential corresponds to about $99.99\%$ of
the expected answer. Furthermore we have calculated the ratio
${|V^{(L,I)}(\xi^*)|}/{T_p}$ for several values of $0\leq\!k^2<1$
at maximal approximation $(L,I)$. We found that this ratio
approaches unity as $k\to 0$, as can be seen in Fig.\ref{fig01}.

\begin{figure}[ht]
\centerline{\includegraphics[{angle=90,height=7.0cm,angle=180,width=8.0cm}]
{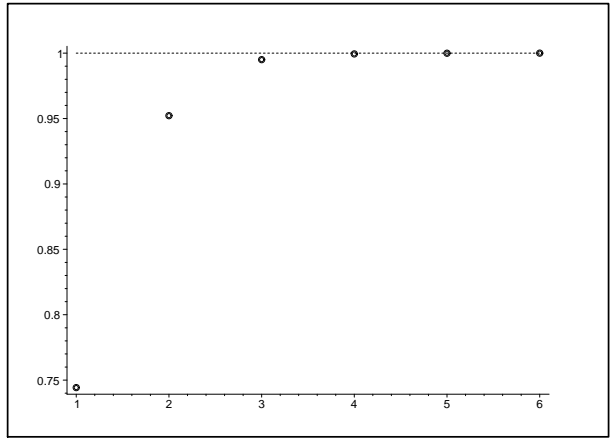}} \caption{The ratio $|V(\xi^*)|/T_p\,$=\,0.744314,\,
0.952164,\, 0.994983,\, 0.999376,\, 0.999927,\,  and\, 0.999974
for the six values $k^2\,$=\,3/4,\, 1/2,\, 1/4,\, 1/8,\, 1/16 and
1/32, respectively, at maximal approximation.}\label{fig01}
\end{figure}

Let us now consider explicit calculations for $k\to0$. It should be
understood that in this limit the sphaleron solution approaches
$\bar{\Phi}(x)\simeq kb\,\sin{bx}$, whose quantum mechanics problem
can still be treated via Lam\'e equation (\ref{lame}). The tachyon
potential obtained in this case is given by \ben \label{k_0th}
V(\xi)&=& { - \xi_0^{2} + 3\,\xi_3 ^{2}+ 3\,\xi_4^{2} + { \frac
{1}{4}} \,{ \frac {\sqrt{2}\,\xi_0^{4}}{\pi }} + { \frac {3}{8}} \,{
\frac {\sqrt{2}\, \xi_1^{4}}{\pi }}+ { \frac {3}{8}} \,{ \frac
{\sqrt{2}\,\xi_2^{4}}{\pi }}+ { \frac {3}{8}} \,{ \frac {
\sqrt{2}\,\xi_3^{4}}{\pi }} + { \frac {3}{8}} \, { \frac
{\sqrt{2}\,\xi_4^{4}}{\pi }}   }\nonumber \\
\mbox{} &&  + { \frac {3}{2}} \, { \frac
{\sqrt{2}\,\xi_4^{2}\,\xi_0^{2}}{\pi }}+ { \frac {3}{4}} \,{ \frac
{\sqrt{2}\,\xi_4^{2} \,\xi_3^{2}}{\pi }}  + { \frac {3}{2}} \,{
\frac {\sqrt{2} \,\xi_3^{2}\,\xi_0 ^{2}}{\pi }}  + { \frac {3}{2}}
\,{ \frac {\sqrt{2}\,\xi_0^{2}\,\xi_1^{2}}{\pi }}  - 6\, { \frac
{\xi_2 \,\xi_1\,\xi_3\,\xi_0}{\pi }}  - 3\,
{ \frac {\xi_2^{2}\,\xi_4\,\xi_0}{\pi }}\nonumber  \\
\mbox{} && + { \frac {3}{2}} \,{ \frac {\sqrt{2}\,
\xi_2^{2}\,\xi_0^{2}}{\pi }} + { \frac {3}{2} } \,{\frac
{\sqrt{2}\,\xi_2^{2}\,\xi_4^{2}}{\pi }}
 + { \frac {3}{2}} \,{ \frac {\sqrt{2}
\,\xi_4^{2}\,\xi_1^{2}}{\pi }}  + { \frac {3}{2}} \, { \frac
{\sqrt{2}\,\xi_2^{2}\,\xi_3^{2}}{\pi }}  \nonumber  \\
\mbox{} && + {\frac {3}{4}} \,{ \frac {\sqrt{2}\,\xi_2^{2}
\,\xi_1^{2}}{\pi }}  + { \frac {3}{2}} \,{ \frac {\sqrt{2}
\,\xi_3^{2}\,\xi_1 ^{2}}{\pi }}   + 3\,{ \frac { \xi_4\,\xi_1^{2}
\,\xi_0}{\pi }}. \een Using (\ref{vev}) we find that the expectation
values $<\xi_1>$, $<\xi_2>$,  $<\xi_3>$, and $<\xi_4>$ are zero.
This means that in the limit $k\to0$ these fields are spurious
states, such that we can remove all of them from the theory. Thus in
this limit the relevant multiscalar tachyon potential simply reads
\ben \label{k_00th} V(\xi)=
-\xi_0^2+\frac{1}{4}\frac{\sqrt{2}}{\pi}\,\xi_0^4, \een whose
nontrivial critical points are $\xi_0^*=\pm\sqrt{\sqrt{2}\pi}$. At
these points the potential assumes the minimal value
$|V(\pm\sqrt{\sqrt{2}\pi})|=(1/2)\sqrt{2}\,\pi$. Here the study of
tachyon condensation restricts only to level zero, i.e., we must
consider just the tachyon field with no further scalars in the
potential.

\begin{figure}[ht]
\centerline{\includegraphics[{angle=90,height=7.0cm,angle=180,width=8.0cm}]
{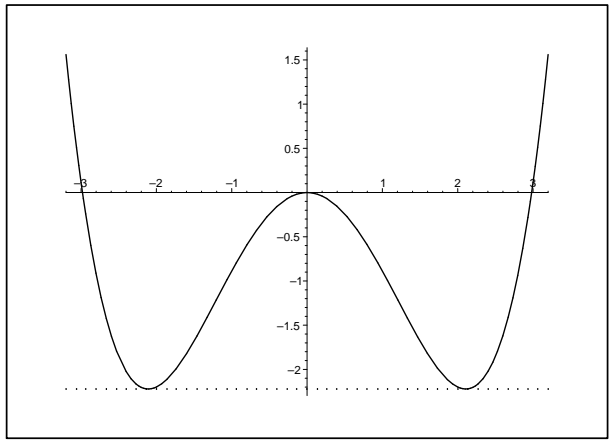}} \caption{The tachyon potential at level zero as $k\to0$. The
dotted line stands for minus the brane tension $T_p$.}\label{fig1}
\end{figure}

Since the $Dp$-brane tension in this case is given exactly by
$T_p=(1/2)\sqrt{2}\,\pi$, thus \ben \label{ratio_0th}
\frac{|V(\xi^*)|}{T_p}=1, \een which corresponds exactly to $100\%$
of the expected answer ! (see Eq.~\ref{conject}) As we mentioned
earlier in the limit $k\to0$ few states are needed to contribute to
the potential. Here we confirm that only the tachyon field $\xi_0$
effectively contributes to the potential even at level zero. The
tachyon potential given by (\ref{k_00th}) behaves as depicted in
Fig.\ref{fig1}. Note that this tachyon has the same mass,
${M_0}^2=-2$, of the tachyon in the original $\Phi^4$ theory given
in (\ref{Vphi}), as it happens in string theory \cite{zwiebach1}.
Note also because the tachyon potential is invariant under $Z_2$
symmetry, it supports a tachyon kink interpolating between the two
minima of the potential at $\xi^*_0=\pm\sqrt{\sqrt{2}\pi}$. This
kink represents a  BPS D-brane of one lower dimension, i.e., a
$D(p-1)$-brane, with tension $T_{p-1}$
\cite{sen10,horava,asad_terashina,berko_sen_zwie,bbn}. We can
readily use the $Dp$-brane action (\ref{eff_actionS}) to obtain the
kink solution $\xi_0(y)$ and its tension $T_{p-1}$ \ben
\xi_0=\pm\xi^*_0\tanh{(y)},\qquad\qquad
T_{p-1}=\frac{4}{3}(\xi^*_0)^2. \een

\section{Conclusions}
\label{conclu}

We investigate the cancelation between the brane tension and the
minimum of the tachyon potential living on the world-volume of a
sphaleron solution representing a non-BPS $Dp$-brane. We consider a
scalar field $\Phi^4$ theory in $p+2$ dimensions with one spatial
coordinate compactified on a circle. This field theory produces a
sphaleron solution that is static and localized in space, but
unstable. In the sphaleron spectrum there is a tachyon mode, the
zero mode and other massive modes. We integrate out these modes over
the compact coordinate to obtain the multiscalar tachyon potential
living on the sphaleron world-volume. The compact coordinate has a
size $L=4K(k)/b$ depending on the elliptic modulus parameter $0\leq
k<1$ of the sphaleron solution. The brane tension $T_p$ is fixed for
a given $k$, while the minimum of the potential $V(\xi^*)$ depends
on the level of approximation. Thus the cancelation between them
depends crucially on the efficiency of the modes in contributing to
the minimum of the potential in the level expansion. In the limit
$k\to1$, the compact coordinate becomes large and the efficiency of
each mode in contributing to the minimum of the potential also
increases. On the other hand, in the limit $k\to0$, this efficiency
for higher modes decreases such that just the tachyon mode
contributes to the minimum of the potential. In this case, we have
shown there exists a perfect cancelation between $T_p$ and
$V(\xi^*)$, which completely agrees with expected answer
(\ref{conject}). One could extend our analysis of tachyon
condensation on sphalerons in field theory to investigations in
string theory. These studies might shed some light on several issues
such as $T$-duality and closed string tachyons.

\acknowledgments

We would like to thank D. Bazeia for discussions,  N. Berkovits for
useful correspondence, and CNPq for partial support.

\end{document}